\documentclass[floatfix, noshowpacs, preprintnumbers, twocolumn, amsmath, amssymb, aps, superscriptaddress, prb]{revtex4}

\pdfoutput=1

\usepackage{graphicx, xcolor}
\usepackage[caption=false]{subfig}
\usepackage{soul}

\newcommand{\ket}[1]{\left| #1 \right>} 
\newcommand{\braket}[2]{\left< #1 \vphantom{#2} \right| \left. #2 \vphantom{#1} \right>}

\begin{document}

\title{Experimental Two-dimensional Quantum Walk on a Photonic Chip}

\author{Hao Tang$^{1,2}$, Xiao-Feng Lin$^{1,2}$, Zhen Feng$^{1,2}$, Jing-Yuan Chen$^{1}$, Jun Gao$^{1,2}$, Ke Sun$^{1}$, Chao-Yue Wang$^{1}$, Peng-Cheng Lai$^{1}$, Xiao-Yun Xu$^{1,2}$, Yao Wang$^{1,2}$, Lu-Feng Qiao$^{1,2}$, Ai-Lin Yang$^{1,2}$ and Xian-Min Jin$^{*}$}

\address{State Key Laboratory of Advanced Optical Communication Systems and Networks, Institute of Natural Sciences $\&$ Department of Physics and Astronomy, Shanghai Jiao Tong University, Shanghai 200240, China}
\address{Synergetic Innovation Center of Quantum Information and Quantum Physics, University of Science and Technology of China, Hefei, Anhui 230026, China}

\email{xianmin.jin@sjtu.edu.cn} 

\maketitle
\textbf{Quantum walks, in virtue of the coherent superposition and quantum interference, possess exponential superiority over its classical counterpart in applications of quantum searching and quantum simulation. The quantum enhanced power is highly related to the state space of quantum walks, which can be expanded by enlarging the photon number and/or the dimensions of the evolution network, but the former is considerably challenging due to probabilistic generation of single photons and multiplicative loss. Here we demonstrate a two-dimensional continuous-time quantum walk by using the external geometry of photonic waveguide arrays, rather than the inner degree of freedoms of photons. Using femtosecond laser direct writing, we construct a large-scale three-dimensional structure which forms a two-dimensional lattice with up to 49$\times$49 nodes on a photonic chip. We demonstrate spatial two-dimensional quantum walks using heralded single photons and single-photon-level imaging. We analyze the quantum transport properties via observing the ballistic evolution pattern and the variance profile, which agree well with simulation results. We further reveal the transient nature that is the unique feature for quantum walks of beyond one dimension. An architecture that allows a walk to freely evolve in all directions and a large scale, combining with defect and disorder control, may bring up powerful and versatile quantum walk machines for classically intractable problems.}

\section{Introduction}
Quantum walks (QWs), the quantum analogue of classical random walks \cite{Aharonov1993, Childs2002}, demonstrate remarkably different behaviours comparing to classical random walks, due to the superposition of the quantum walker in its paths. This very distinct feature leads the quantum walks to be a stunningly powerful approach to quantum information algorithms\cite{Ambainis2003,Shenvi2003, Childs2004, Sanchez2012, Schuld2014}, and quantum simulation for various systems \cite{Mulken2011, Lambert2013, Aspuru2012}. For instance, theoretical research has revealed that QWs propagating in one dimension (1D) possess superior transport properties to 1D classical random walks\cite{Mulken2006}, and the coherence in QWs is crucial in simulating energy transport in the photosynthetic process\cite{Mulken2011, Lambert2013}. The potential of applying QWs in machine learning algorithms such as artificial neural network \cite{Schuld2014} also draw wide attention from multidisciplinary researchers. Inspired by the prospects of QWs, many endeavours have been made to realize QWs in different physics systems, including nuclear magnetic resonance\cite{Du2003}, trapped neutral atoms\cite{Karski2009}, trapped ions\cite{Schmitz2009}, and photonic systems\cite{Peruzzo2010,Biggerstaff2016,Perets2008}. 

\begin{figure*}[ht!]
\includegraphics[width=0.8\textwidth]{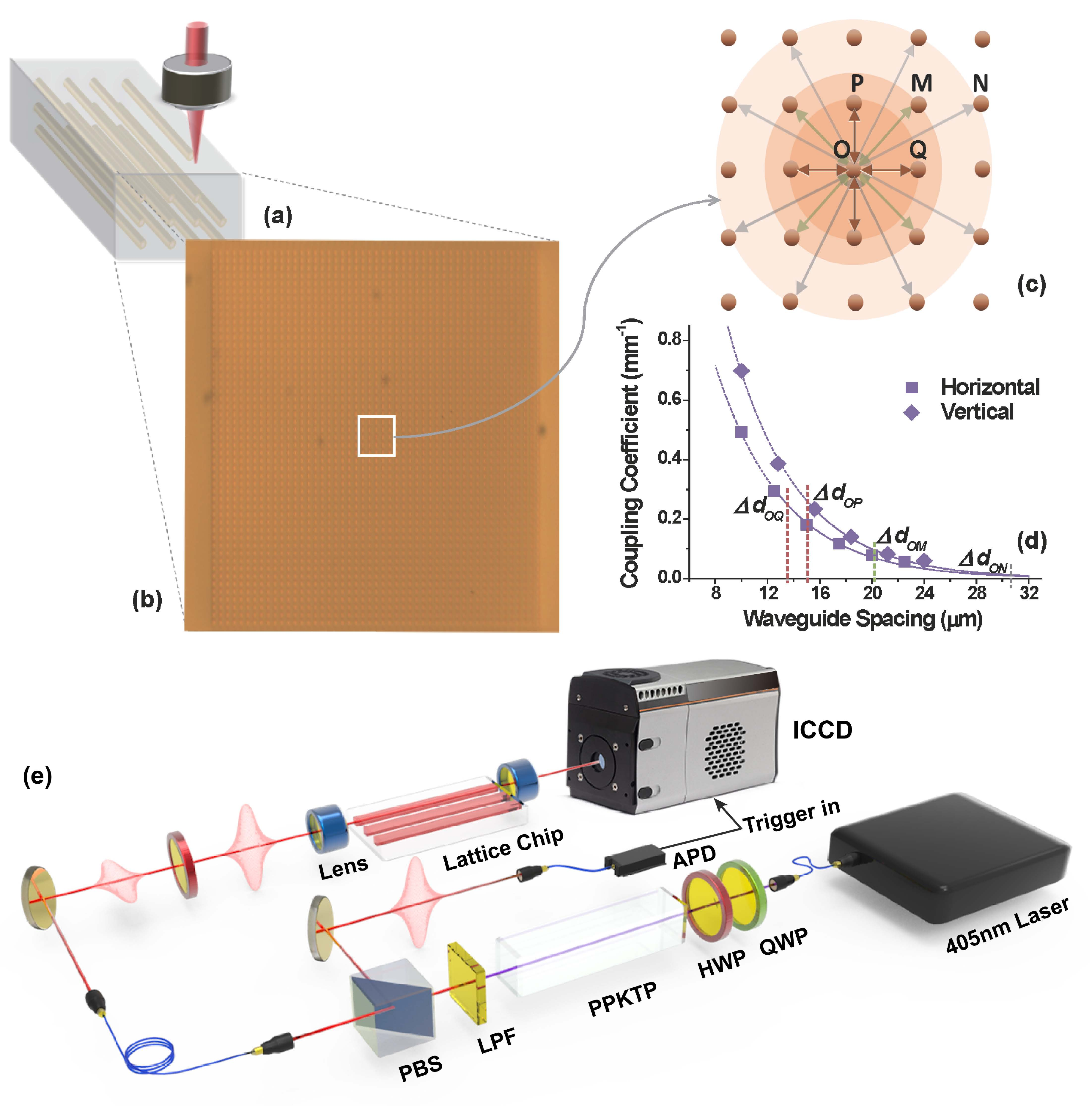}
\caption{\textbf{Experimental layout.} ({\bf a}) Schematic diagram of 3D waveguide arrays fabrication using the femtosecond laser direct writing technique. ({\bf b}) Photographed cross section of a photonic lattice studied in this experiment. ({\bf c}) Schematic diagram of one waveguide coupling to other waveguides in the 3D waveguide arrays, and ({\bf d}) the corresponding coupling coefficient $C$ for different centre-to-centre waveguide spacings in horizontal and vertical directions. ({\bf e}) Setup of single photon experiment. Each photonic chip to be tested incorporates many sets of 3D waveguide arrays. ICCD: intensified charge coupled device, APD: avalanched photo diode, PBS: polarized beam splitter, HWP: half wave plate, QWP: quarter wave plate, LPF: long pass filter, PPKTP: periodically poled KTP crystal.}
\label{fig:QFTConcept}
\end{figure*} 

However, these experimental implementations reveal a very evident limitation, that the realized quantum walk is normally of only one dimension, and  the evolving scale of QWs remains very small. Simple demonstration of 1D QW could not suffice the ever growing demand for further speed-up of certain quantum algorithms, or the simulation of quantum systems of a much higher complexity\cite{Poulios2014, Gao2016}. In the spatial search algorithm, a quantum walk outperforms its classical counterparts only when the dimension is higher than one\cite{Tulsi2008}; In the simulation of graphene, photosynthesis, or neural network systems, these complex networks always intuitively have high dimensions. Experimental research on quantum walks of beyond 1D becomes indispensable, and a few attempts having covered 2D QWs in experiments are worth noted. A discrete-time 2D quantum walk was achieved in the fibre network system by dynamically controlling the time interval of two walkers\cite{Schreiber2012}, in the so-called delayed-choice scheme \cite{Jeong2013}, or using two walkers sharing coins \cite{Xue2015}. They ingeniously use either time-polarization dimension or the analog from two walkers acting on 1D graph to represent one walker on a 2D lattice, and the 2D lattice does not physically occur.  A quasi-2D continuous-time quantum walk was explored in the waveguide coupled in a `Swiss cross' arrangement\cite{Poulios2014}, but this is not, strictly speaking, a 2D quantum walk, because photons could not freely propagate in the diagonal and many other directions as they suppose to do in the 2D array. 

In this paper, we for the first time experimentally observe the evolution of 2D continuous-time quantum walks with single photons on the 2D waveguide array. We set up the heralded single-photon source and measure the evolution results that agree well with theoretical simulation using an ultra-low-noise single-photon-level imaging technique. We further analyze the transport and recurrent properties, measured from the variance and the probability from the initial waveguide, respectively. We experimentally verify the unique features for two-dimensional quantum walks that differ from both classical random walks and quantum walks of one-dimension. 

\section{Main}
Photons propagating through the coupled waveguide arrays can be described by the Hamiltonian:
$$H=\sum_{i}^N \beta_i a_i^\dagger a_i + \sum_{i \neq j}^N C_{i,j} a_i^\dagger a_j\eqno{(1)}$$
where $\beta_i$ is propagating constant in waveguide $i$, $C_{i,j}$ is the coupling strength between waveguide $i$ and $j$. For a uniform array, all $\beta_i$ is regarded equal to $\beta$, and $C_{i,j}$ that mainly depends on waveguide spacing can be obtained via a coupled mode approach \cite{Szameit2007}. 

In our implementation, we fabricate two-dimensional waveguide arrays using femtosecond laser writing techniques \cite{Davis1996} (Fig.1.a). The waveguides are written in different depths of the borosilicate glass to form a two-dimensional array\cite{Thomson2011} from the cross-section view (Fig.1.b). The centre-to-centre spacing between two nearest waveguides is set as a spacing unit that is 15 $\mu$m in the vertical direction ($\Delta d_V$) and 13.5 $\mu$m in the horizontal direction ($\Delta d_H$). In such a two-dimensional array, each waveguide is involved into comprehensive coupling with surrounded waveguides, e.g., Waveguide $O$ has different waveguide spacings to Waveguide $P$, $Q$, $M$ and $N$ as marked in Fig.1.c, namely, $\Delta d_V$, $\Delta d_H$, $\sqrt{\Delta d_H^2+\Delta d_V^2}$ and $\sqrt{(2\Delta d_H)^2+\Delta d_V^2}$ for $\Delta d_{PO}$, $\Delta d_{QO}$, $\Delta d_{MO}$ and $\Delta d_{NO}$ respectively.  Such differences in waveguide spacings and waveguide-pair orientations affect the coupling coefficient significantly, as shown in Fig.1.d. Through the measured value of $C$ following the standard method \cite{Szameit2007}, we observe the exponential decay as waveguide spacing increases and some discrepancy of $C$ in different directions. We hence select $\Delta d_H$ and $\Delta d_V$ to ensure uniform coupling coefficients for nearest waveguide pairs in the horizontal and vertical directions. For other waveguide pairs in inclined directions, such as Pair $M$-$O$ and Pair $N$-$O$ in Fig.1.c, as the directional discrepancy of $C$ gets smaller when waveguide spacing increases, we use the average of the horizontal and vertical value at the corresponding spacing for their coupling coefficient. 

For a quantum walk that evolves along the waveguide, the propagation length $z$ is proportional to the propagation time by $z = ct$, where $c$ is the speed of light in the waveguide, and hence all terms that are a function of $t$ would use $z$ instead in this paper for simplicity. The wavefunction that evolves from an initial wavefunction satisfies: 
$$\ket{\Psi(z)}=e^{-iHz}\ket{\Psi(0)}\eqno{(2)}$$  
where $\ket{\Psi(z)}=\sum_ja_j(z)\ket{j}$, and $|a_j(z)|^2=|\braket{j|\Psi(z)}|^2=P_j(z)$
respectively. $|a_j(z)|^2$ and $P_j(z)$ is the probability of the walker\cite{Izaac2015} being found at waveguide $j$ at the propagation length $z$. As is shown in Fig.1.e, we observe the dynamics by injecting a vertically polarized heralded single photon source (810 nm) into the central waveguide\cite{Fedrizzi2007} and measuring the evolution patterns using an ICCD camera. More details about our single-photon source and the ultra-low-noise single-photon-level imaging can be found in the Method section. 

\begin{figure}[t!]
\includegraphics[width=0.43\textwidth]{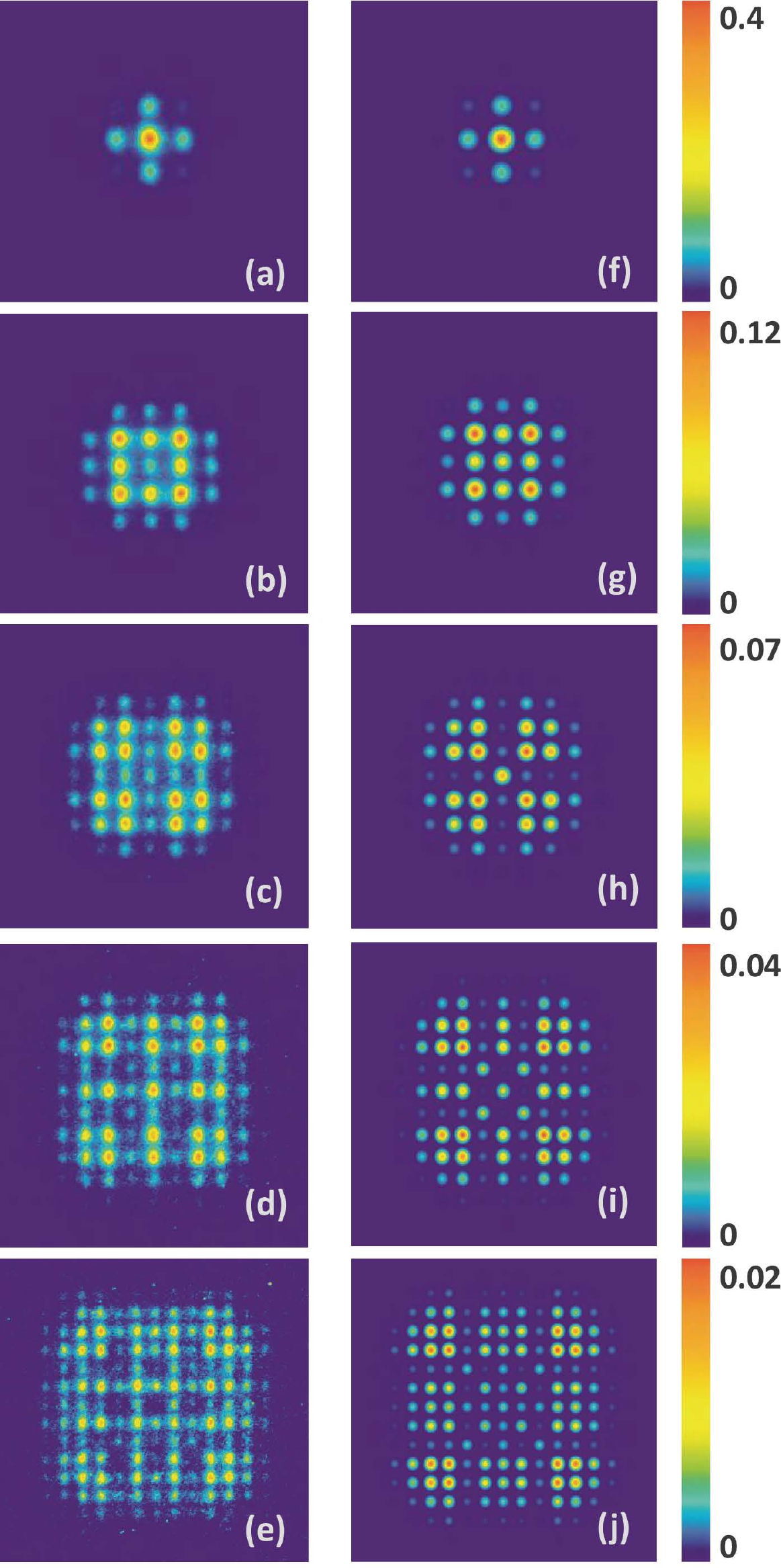}
\caption{\textbf{2D quantum walks of different propagation lengths.} ({\bf a-e}) Experimentally obtained probability distribution of heralded single photons and ({\bf f-j}) theoretical probability distribution. The propagation lengths are: 1.81 mm for ({\bf a}) and ({\bf f}), 3.31 mm for ({\bf b}) and ({\bf g}), 4.81 mm for ({\bf c}) and ({\bf h}), 7.31 mm for ({\bf d}) and ({\bf i}), and 9.81 mm for ({\bf e}) and ({\bf j}). 
}
\label{fig:strutturaChip}
\end{figure}

These two-dimensional patterns of different propagation lengths from both experimental evolution of heralded single photons and theoretical simulations are then collected (Fig.2). Clearly, the intensity peaks always emerge at the diagonal positions, and they move further in these directions when the propagation length $z$ increases. The similarity between two probability distributions $\Gamma_{i,j}$ and $\Gamma_{i,j}'$ can be defined by\cite{Peruzzo2010}: $S=(\sum_{i,j}(\Gamma_{i,j} \Gamma_{i,j}')^{1/2})^2/\sum_{i,j}\Gamma_{i,j} \sum_{i,j} \Gamma_{i,j}'$. For the five pairs in Fig. 2, the similarities are calculated as 0.961  (a \& f), 0.957 (b \& g), 0.920 (c \& h), 0.917 (d \& i) and 0.913 (e \& j), respectively. Therefore, there is a good match between experimental evolution patterns and the theoretical results of two-dimensional quantum walks.

\begin{figure*}[ht!]
\includegraphics[width=1\textwidth]{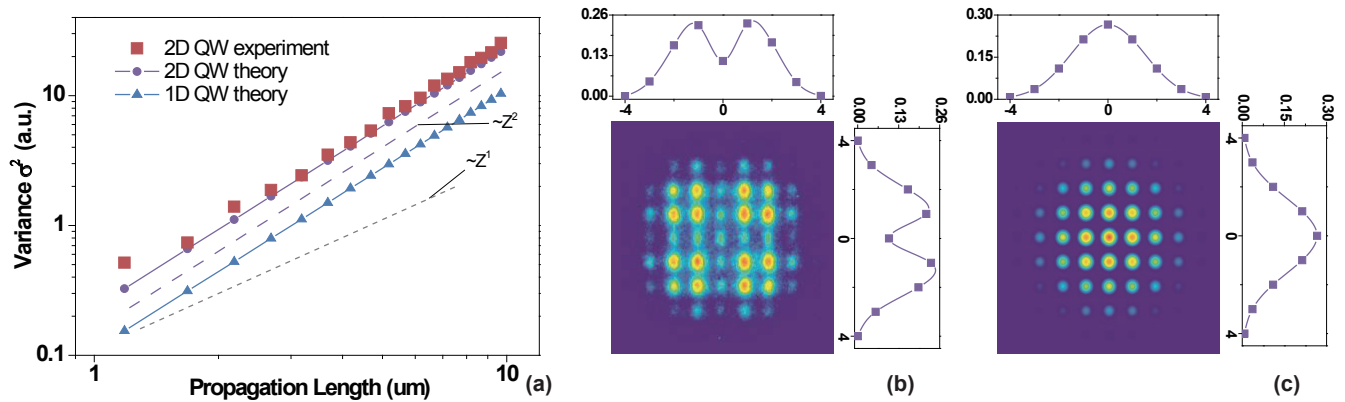}
\caption{\textbf{ The transport properties of quantum walks.} ({\bf a}) The variance against propagation length for experimental 2D QWs, theoretical 2D QWs and theoretical 1D QWs. ({\bf b}) An evolution pattern of a 2D quantum walk from heralded-single-photon experiment at a propagation length $z$=4.31 mm and its projection profile onto $x$ and $y$ axis. ({\bf c}) A theoretical evolution pattern of a 2D classical random walk in a 2D Gaussian distribution with a sigma of 1.5 spacing units and its projection profile onto $x$ and $y$ axis. }
\label{fig:apparato}
\end{figure*}

\subsection{The transport properties of quantum walks}

We know quantum walks have unique transport properties, which could be examined from the variance against the propagation length, as defined in Eq.(3):
 $$\sigma(z)^2=\frac{\sum_{i=1}^N\Delta l_i^2P_i(z)}{\sum_{i=1}^NP_i(z)}\eqno{(3)}$$
where $\Delta l_i$ is the normalized spacing between waveguide $i$ and the central waveguide where the single photons are injected into. Plotting the variance-propagation length relationship with double-logarithmic axes, the ballistic 1D quantum walk is known for yielding a straight line with slope 2, while the diffusive 1D classical random walk results in a straight line with slope 1, i.e., QW transports quadratically faster than the classical random walk \cite{Eichelkraut2013}.

The variance for both one-dimensional quantum walks in theory, and two-dimensional quantum walks in theory and in experiments are presented in Fig.3.a. All quantum walks have the same coupling coefficient for waveguide pairs of the nearest spacing, and the walks evolve in a lattice large enough to ignore boundary effects. For two-dimensional quantum walk, the experimental results agree well with the theoretical ones. The variance from one-dimensional quantum walk in theory goes all the way below the two-dimensional case, as a walker can move in more directions in the latter. However, the variance for all these quantum walks follows the trend of slope 2 rather than slope 1, suggesting the universal ballistic spreading for both one and two-dimensional quantum walks, which distinguishes them from diffusive classical random walks. 

Projecting the evolution patterns of a 2D quantum walk and a 2D classical random walk onto $x$ axis and $y$ axis (Fig 3.b and c), the random walk in a two-dimensional Gaussian distribution \cite{Manouchehri2007} has the projection profiles of a 1D Gaussian distribution, while the projection profiles for the quantum case show a ballistic shape similar to the 1D quantum walks. It indicates that the intensity peaks in random walks always remain in the centre, but those in quantum walks always move to all frontiers, causing a larger variance for the latter. 
 
\subsection{The recurrent properties of quantum walks}

We further investigate the difference between quantum walks of different dimensions, which can be gauged by $P_0(z)$ and P$\rm \acute{o}$lya number, two indices that concern the recurrent properties of a walker in a network\cite{Darazs2010, Darazs2014}.

$P_0(z)$, the probability of a walker being found at the initial waveguide after a propagation length $z$, is plotted in Fig.4.a. All quantum walks have a decreasing $P_0(z)$ as $z$ increases, but follow different asymptotic lines. A walker in a 2D lattice evolves away from the original site much faster through many additional paths and is less likely to move back (with a smaller oscillation) comparing to the 1D scenario.  

A system can be judged to be recurrent or transient depending on the P$\rm \acute{o}$lya number, through the definition\cite{Darazs2010, Darazs2014}: 
$$P=1-\prod_{m=1}^\infty[1-P_0(z_m)]\eqno{(4)}$$   
where $z_m$ is a set of propagation lengths sampled periodically \cite{Darazs2010}. When the P$\rm \acute{o}$lya number is 1, a system is recurrent, because $P_0(z_m)$ can always be a large value to make $\prod_{i=1}^\infty[1-P_0(z_m)]$ close to zero, while for a transient system, $P_0(z_m)$ quickly drops to a very marginal value so the P$\rm \acute{o}$lya number would be smaller than one \cite{Stefanak2008, Kollar2010}.

\begin{figure*}[ht!]
\includegraphics[width=0.9\textwidth]{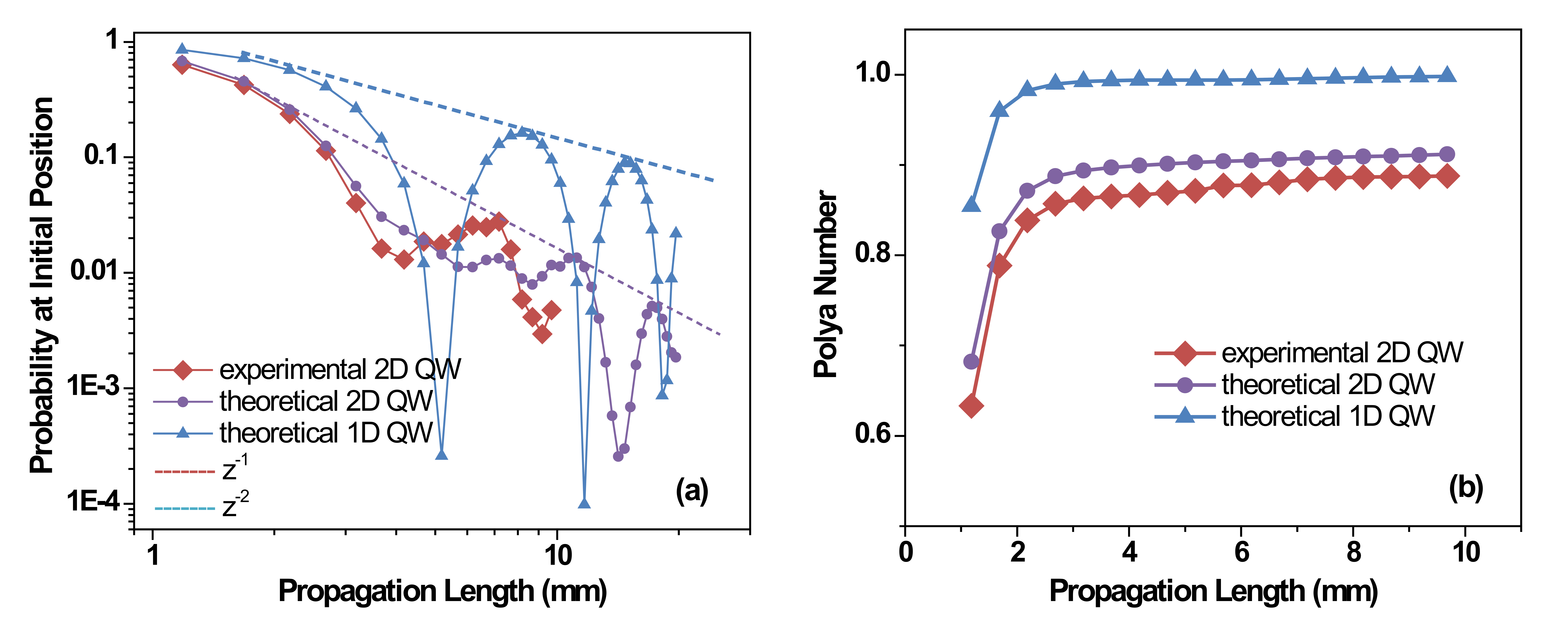}
\caption{\textbf{The recurrent properties.} ({\bf a}) Probability at the initial position against propagation length and ({\bf b}) P$\rm \acute{o}$lya number against propagation length for experimental 2D QWs, theoretical 2D QWs and theoretical 1D QWs.}
\label{fig:Results4}
\end{figure*}

Two-dimensional QWs in experiment and in theory, and one-dimensional QWs in theory show a P$\rm \acute{o}$lya number approaching 0.887, 0.912 and 0.998, respectively (Fig.4.b). Clearly, the 2D QW is much less inclined to be recurrent than 1D QW. Further interpretation\cite{Darazs2010} comes from the asymptotic features $z^{-d}$. It has been pointed out that transient systems tend to have a value of $d$ larger than 1, while $d$ for recurrent systems would be equal to or go below 1. From Fig.4.a, the 2D quantum walks in experiment and in theory both follow an asymptotic line $z^{-2}$, revealing the transient nature for these 2D continuous-time quantum walks in our implementation. We for the first time measure the transient nature of a 2D quantum walk in experiment, which makes it different from all experimentally realized quantum walks that were either in 1D or in 2D with limited scales. 

\section{Discussion}

Here, we have demonstrated strong capacity in achieving large-scale three-dimensional photonic chips and ultra-low-noise single-photon-level imaging techniques that are crucial for the implementation and measurement of our spatial two-dimensional continuous-time quantum walks. The first and large-scale realization of real spatial 2D quantum walk may not only be fundamentally interesting but also provide a powerful platform for quantum simulation and quantum computing. Since we increase the dimensions by the evolution network geometry, even with single walker, photon evolution on lattices up to 49$\times$49 nodes may lead to a huge state space being large enough to explore new physics in entirely new regimes. Quantum advantage/supremacy may also be explored in such platform using analog quantum computing protocols, such as 2D Boson sampling\cite{Spring2012}, fast hitting\cite{Childs2002} and even universal quantum computing protocols\cite{Childs2013}, instead of using circuit-model protocols of universal quantum computing.

The spatial structure itself can also be freely fabricated with special geometric arrangement, defect, disorder, topological structure in a programmable way, which may offer a new approach of Hamiltonian engineering to enable designing and building quantum simulators on demand on a photonic chip. Such a Hamiltonian engineering can be realized by adding waveguide curvature, variation of the fabrication power or dynamic waveguide spacings, etc. Through these we could potentially extend the issue of localization in quantum walks to higher dimensions\cite{Lahini2008}, as well as exploring topological photonics and the simulation of quantum open systems in photonic lattices\cite{Caruso2016}. 

Further, we would go beyond two dimensions through various ways. Quantum simulations in (2+1) dimensions are possible and their dynamic properties can be explored if we introduce time-varying Hamiltonian along the propagating axis. For issues such as quantum walks in bosonic and fermionic behaviours \cite{Sansoni2012}, multi-particle entanglement and evolution, etc., the multi-photon source interfaced to the robust and precise photonic chips could give the research of high-dimension quantum systems an instant boost, and demonstrate its strong potential for quantum simulation in a highly complex regime.

\section*{Methods}
 
\textbf{Photonic lattice preparation:} Waveguide arrays were prepared by steering a femtosecond laser (10W, 1026~nm, 290~fs pulse duration, 1~MHz repetition rate and 513~nm working frequency) into an spatial light modulator (SLM) to create burst trains onto a borosilicate substrate with a 50X objective lens (numerical aperture:~0.55) at a constant velocity of 10~mm/s. Power and SLM compensation were processed to ensure the waveguides to be uniform and depth independent\cite{Feng2016}. The borosilicate glass wafer is of a size 1$\times$20$\times$20~mm, and consists of 20 set of lattices of different evolution lengths from 0.31~mm to 9.81~mm. Each lattices has 49$\times$49 waveguides in a size of 0.72~mm$\times$0.648~mm in the cross section view.

\textbf{Single-photon source and imaging:} A 405nm diode laser pumped a PPKTP crystal to generate pairs of 810nm via type II spontaneous parametric downconversion. The resulted single-channel count rate and two-channel coincidence count rate reach 510000 and 120000, respectively. The generated photon pairs then pass a 810nm band-pass filter and a polarized beam splitter to be divided to two purified components of horizontal and vertical polarization. The vertically polarized photon was coupled into a single-mode optical fiber and then injected into the photonic chips, while the horizontally polarized photon is connected to a single photon detector that sets a trigger for heralding the vertically polarized photons on ICCD camera with a time slot of 10 ns. If without such an external trigger, the measured patterns would come from light in thermal states rather than single-photon states. ICCD camera captures each evolution pattern from the photon output end of the photonic chip, after accumulating single photon injections in the `external' mode for around an hour. 

\textbf{Simulation of light field evolution:} Solving Eq.(2) requires a matrix exponential method and this yields the light evolving pattern that contains the probability matrix for all waveguides. The Pade approximation function \cite{Moler1978} in Matlab is used in the simulation. The calculated probability for each waveguide is then treated to be a Gaussian spot with the spot intensity proportional to the probability, in order to visualize the comparison between the theoretical and experimental patterns.

\bigskip

\textbf{Acknowledgements.} 
The authors thank J.-W. Pan for helpful discussions. This research is supported by the National Key Research and Development Program of China (2017YFA0303700), National Natural Science Foundation of China (Grant No. 61734005, 11761141014, 11690033, 11374211), the Innovation Program of Shanghai Municipal Education Commission, Shanghai Science and Technology Development Funds, and the open fund from HPCL (No. 201511-01), X.-M.J. acknowledges support from the National Young 1000 Talents Plan.
\end{document}